\documentstyle[times,pramana,epsf,floats]{ias}

\newcommand{\lton}{\mathrel{\lower.9ex
                  \hbox{$\stackrel{\displaystyle <}{\sim}$}}}

\begin{document}
\mark{{QGP Theory: Status and Perspectives}{Steffen A. Bass}}
\title{QGP Theory: Status and Perspectives}

\author{Steffen A. Bass}
\address{Department of Physics, Duke University\\
        \& RIKEN-BNL Research Center, Brookhaven National Laboratory}
%\pacs{2.0}
\abstract{The current status of Quark-Gluon-Plasma Theory is reviewed.
Special emphasis is placed on QGP signatures, 
the interpretation of current data and
what to expect from RHIC in the near future.
}

\maketitle
\section{Introduction}

Ultra-relativistic heavy ion collisions 
offer the unique opportunity to probe
highly excited dense nuclear matter under controlled laboratory
conditions. 
One of the main  driving forces for these studies is the expectation that an
entirely new form of matter may be created in such reactions. This
form of matter, called the Quark Gluon Plasma (QGP), is the QCD
analogue of the plasma phase of ordinary atomic matter. However, unlike such
ordinary plasmas, the deconfined quanta of a QGP are not directly
observable because of the fundamental confining property of the
physical QCD vacuum.
What is observable are hadronic and
leptonic residues of the transient QGP state.  
The QGP state formed in nuclear collisions is a
transient rearrangement of the correlations among quarks and 
gluons contained in the incident baryons  into a larger but globally still
color neutral system with however remarkable theoretical properties,
such as restored chiral symmetry and deconfinement.
The task with heavy ion reactions is to provide experimental information
on this fundamental prediction of the Standard Model (for recent
reviews on QGP signatures, please see \cite{qgpreviews}).

\section{The QCD Phase Diagram}

QCD is a non-abelian gauge theory,
it's basic constituents are quarks and anti-quarks interacting
through the exchange of color-charged gluons. 
At very high temperatures and densities, in the domain of weak 
coupling between quarks and gluons, long range interactions are dynamically 
screened \cite{collins75a,polyakov77a}.
Quarks and gluons are then no longer confined to bound hadronic states
(``deconfinement''). Furthermore, chiral symmetry is restored -- for
baryon-free matter -- apparently at the same temperature $T_C$.

\begin{figure}
 \centerline{\epsfxsize=0.6\textwidth\epsfbox{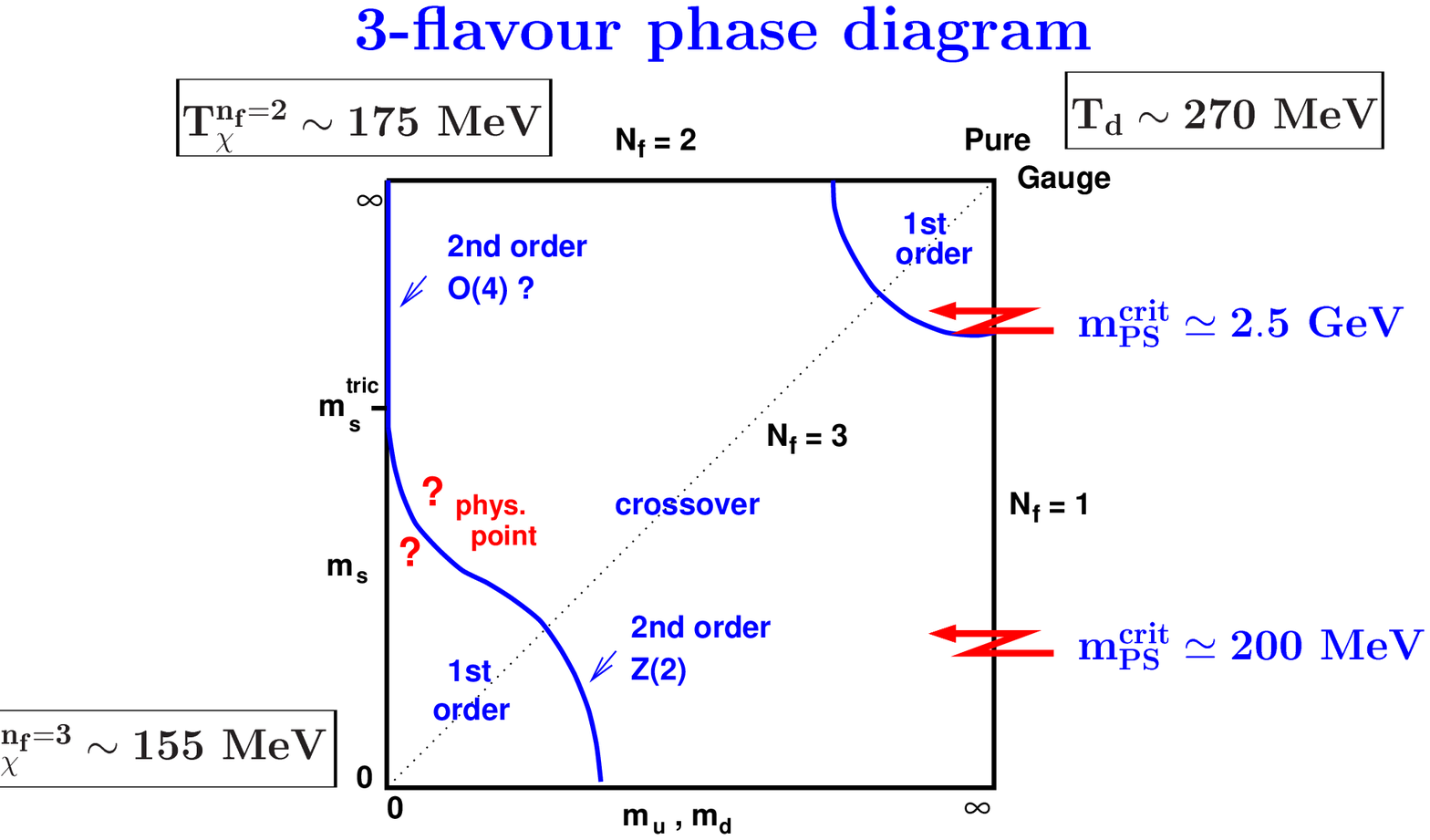} 
\hfill \epsfxsize=0.4\textwidth\epsfbox{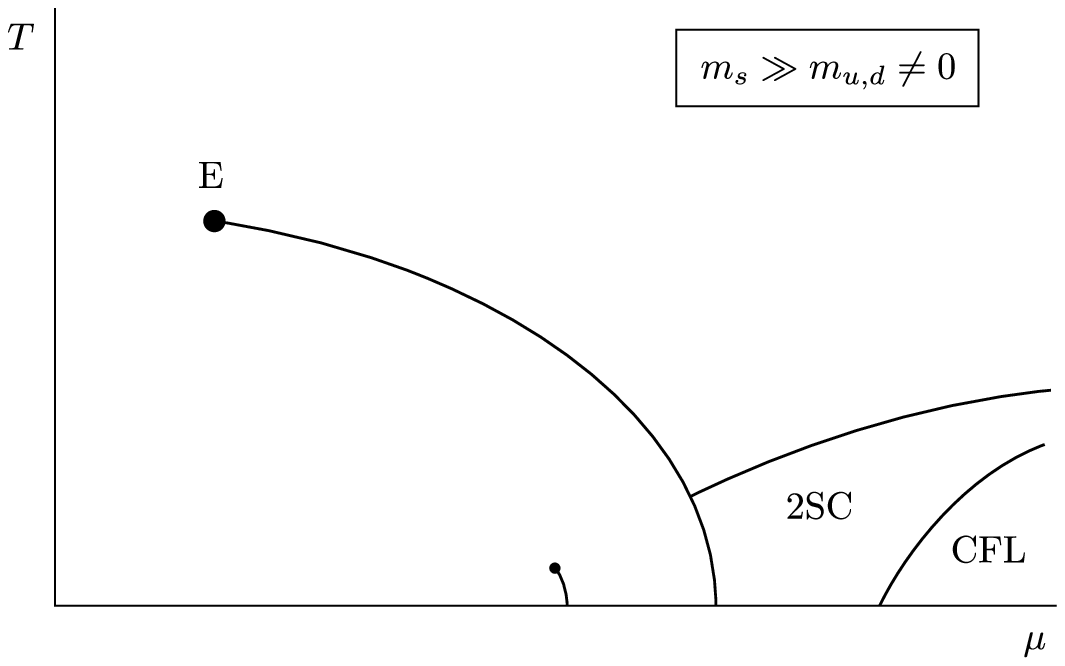}
}
\caption{\label{latticefig1} Left frame: QCD phase diagram of 3-flavor
QCD with degenerate (u,d)-quark masses and a strange quark mass $m_s$. 
Right frame: schematic QCD phase diagram
as a function of $\mu_B$.
}
\end{figure} 

Lattice gauge simulations of QCD \cite{wilson74a,creutz83a}, 
provide the only rigorous method to compute
the {\em equation of state} of strongly
interacting elementary particle matter. In principle
both, the non-perturbative hadronic matter 
and the non-perturbative QGP phases of QCD can be investigated.
The main disadvantage of lattice simulations is the current 
practical restriction to
finite, periodic, baryon free systems in global equilibrium, a scenario far
away from the highly inhomogeneous off-equilibrium
situation found in complex heavy-ion reactions.

Lattice calculations
yield a critical temperature of $T_C \approx 270$ MeV
in the quenched approximation \cite{laermann96a,karsch}
-- where neither dynamical quarks, nor a
chiral phase transition exist. Simulations including
dynamical quarks at $\mu_B=0$ indicate a critical temperature in the order of
$T_C \approx 165$ MeV (see figure \ref{latticefig1}, 
taken from \cite{karsch}). 

Many striking QGP signatures 
depend heavily on the assumption of
a first order phase transition and the existence of a mixed phase of 
QCD matter. However, both, the order of the phase transition as well
as the critical temperature, depend on the parameters of the calculation,
namely the number of quark flavors and their masses.
For the most realistic case of QCD with two flavors of light quarks 
with masses between 5 and 10 MeV and one flavor with a mass around
200 MeV,
the situation remains unclear: 
the order of the phase transition seems to depend on the numerical 
values for the masses of the light and heavy quarks \cite{brownfr90a,karsch}. 
If the latter is too heavy, the transition might be smeared out to a mere
rapid increase of the energy density over a small temperature interval.
In this case the use of simple deconfinement scenarios 
may lead to wrong expectations for observables.
The elementary excitations in such a phase transition region 
ought not be described by quarks and gluons but could 
physically resemble more 
hadronic excitations with strongly modified ``in--medium'' properties
\cite{detar88a}.

This raises a practical question, whether 
conclusions based on $\mu_B=0$ estimates, might misguide physical
argumentation for observables in nuclear collisions.
This warning is particularly appropriate for those QGP-signals, where a 50\%
quantitative change of an observable is used to differentiate 
QGP production scenarios from ordinary hadronic transport ones.

The inclusion of the second most important thermodynamic variable, the chemical
potential $\mu_B$ into a full fletched lattice-QCD calculation has
recently been attempted \cite{fodor}, but still carries large systematical
uncertainties. These calculations predict the existence of a tricritical
point -- only for $\mu_B > \mu_{B,crit}$ would a phase-boundary with a 
possibly 1st order phase-transition exist.

An additional complication is that for systems of finite volume
(V$\le$ 125 fm$^3$) the deconfinement cannot be complete. Fluctuations
lead to a finite probability of the hadronic phase above $T_C$.
The sharp discontinuity (e.g. $\varepsilon/T^4$) is thus smeared out
\cite{spieles97b}. 

Recently, a lot of progress has been made towards understanding the QCD phase
diagram at large chemical potential $\mu_B$ \cite{csc}. At 
very high densities, asymptotic freedom and BCS theory have been utilized 
to predict new, superconducting, phases of QCD matter.
Depending on the mass of the strange quark ($m_s \gg m_{u,d}$), first a
two flavor color-superconductor ($m_s \to \infty$) or a color-flavor locked
superconducting phase ($m_s$ decreasing with $\mu_B$) are found
(see right frame of figure~\ref{latticefig1}, taken from \cite{rajagopal}).
The color-flavor locking refers to the symmetry of the ground state only
being given under a simultaneous transformation of color and flavor.
It should be noted, however, that these new phases
of high-density QCD will most likely not be found in heavy-ion 
reactions (the temperature in such collisions is too high for color
superconductivity to occur), but may rather have a large impact in our
understanding of matter in neutron stars and other cold super-dense 
objects (for a review on the topic, please see \cite{rajagopal}).

\section{Dynamics: Transport Theory at SPS and RHIC}

\begin{figure}[t]
\epsfxsize=12cm
\centerline{\epsfbox{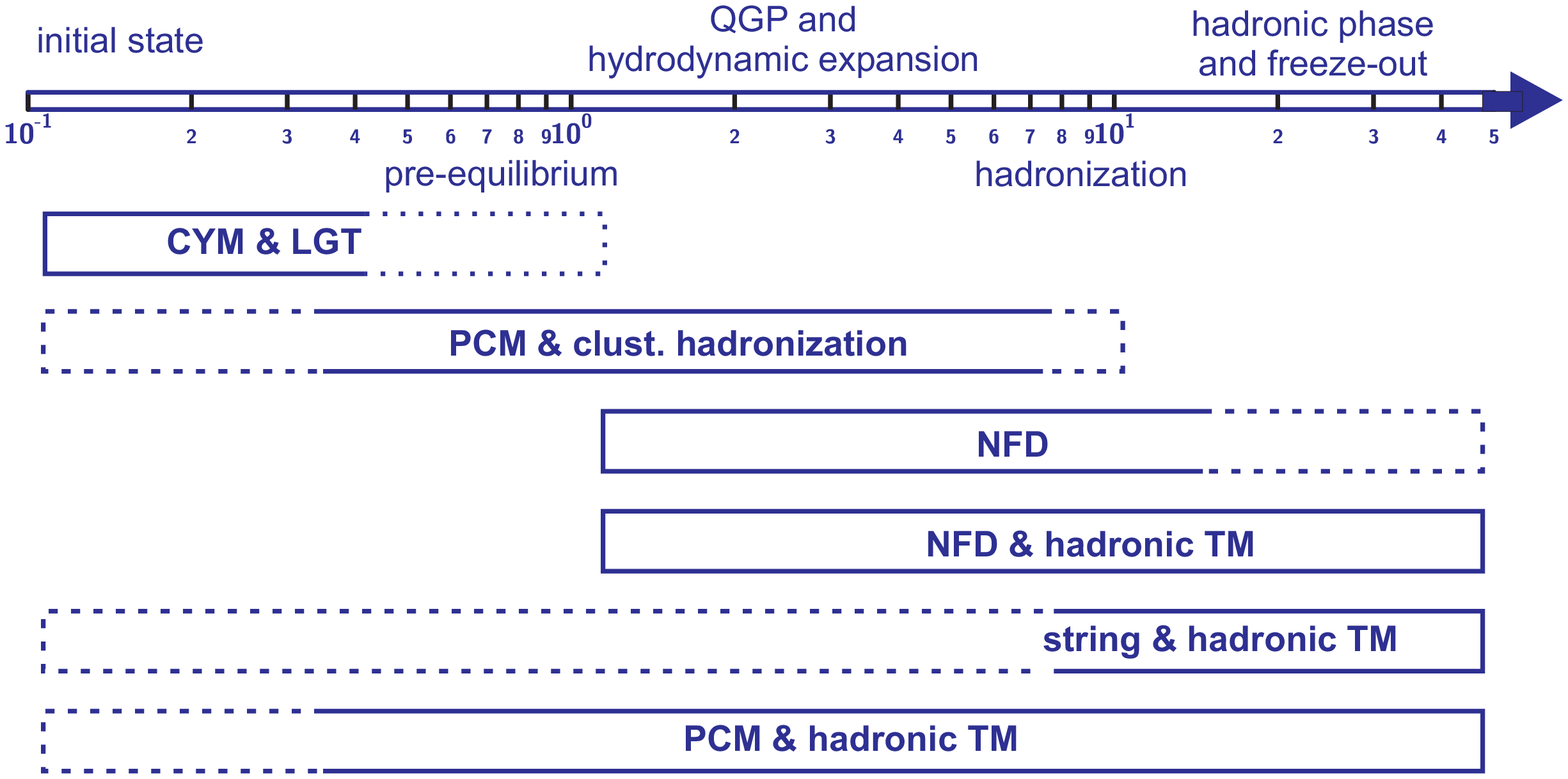}}
%\vspace{-10mm}
\caption{\label{ttover}
Transport theory approaches for RHIC and their
range of applicability.}
\end{figure}

Transport Theory offers the unique capability of connecting experimentally
observable quantities in a relativistic heavy ion collision with its
time evolution and reaction dynamics, thus giving crucial insights
into the possible formation of a transient deconfined phase of
hot and dense matter, the Quark-Gluon-Plasma.

Figure~\ref{ttover} provides an overview
of different transport theoretical ansatzes currently in the literature
 for the description of a relativistic heavy ion collision at RHIC
energies. The timeline shows a {\em best case scenario} for what to expect: 
the formation of a thermalized QGP with 
subsequent hadronization and freeze-out.
Bands with solid lines denote the safe range of applicability
for the respective transport approach, whereas dashed/dotted bands 
refer to areas in which the approach is still applied but where the
assumptions on which the approach is based upon may be questionable
or not valid anymore.

The {\em initial state} and early {\em pre-equilibrium  phase} are best described
in Classical Yang-Mills theory (CYM) \cite{MLV}
or Lattice Gauge Transport (LGT) \cite{BMP}
calculations - only these classes of models treat the coherence of the
initial state correctly, but do not provide any meaningful dynamics for
the later reaction stages.

The Parton Cascade Model (PCM) \cite{PCM} and related pQCD approaches
\cite{hijing}
treat the initial state as incoherent parton configuration,
but are very well suited for the {\em pre-equilibrium phase} and subsequent
thermalization, leading to a {\em QGP and hydrodynamic expansion}. 
Microscopic degrees of freedom in this ansatz are quarks and gluons
which are propagated according to a Boltzmann Equation with a collision
term using leading order pQCD cross sections.
Augmented
with a cluster hadronization ansatz the PCM is applicable up
to {\em hadronization}. The range of this model can be even further extended
if it is combined with a hadronic cascade which treats the {\em hadronic phase
and freeze-out} correctly \cite{bass_vni}.

Nuclear Fluid Dynamics (NFD, see e.g. \cite{Bj,DumRi})
is ideally suited for the {\em QGP and hydrodynamic expansion} reaction phase,
but breaks down in the later, dilute, stages of the reaction when the
mean free paths of the hadrons become large and flavor degrees of freedom
are important. The reach of NFD can also be extended by combining it
with a microscopic hadronic cascade model -- this kind of hybrid approach
(dubbed {\em hydro plus micro}) was pioneered in \cite{hu_main} 
and has been now also taken up by other groups \cite{Teaney:2000cw}. 
It is to date the most successful approach
for describing the soft physics at RHIC. The biggest advantage of NFD is
that it directly incorporates an equation of state as input - one of it's
largest disadvantages is that it requires thermalized initial conditions
and one is not able to do an ab-initio calculation.

Last but not least,  string and
hadronic transport models \cite{rqmd,urqmd} 
have been very successful in the AGS and SPS energy domains.
They treat the early reaction phase as a superposition of hadron-hadron
strings and are thus ill suited to describe the microscopic reaction dynamics
of deconfined degrees of freedom. In the later reaction stages, however,
they are the best suited approach, 
since they incorporate the full spectrum of degrees of freedom of a hadron
gas (including flavor dependent cross sections).

However, it is important to note that there is not a single transport 
theoretical ansatz currently available, which is able to
cover the entire time-evolution of a collision at RHIC in one self-consistent
approach.

\section{SPS: highlights and open questions}

During the  last eight years the heavy ion research  at the CERN-SPS
has succeeded to achieve the measurement
of a wide spectrum of observables relevant for a QGP search. Table~\ref{tab1}
provides an overview of these measurements together with an assessment
if the results would be compatible with the possible creation of a new 
form of deconfined matter \cite{cern_announcement,heinz}.
In the following sections, several of these results will be highlighted
in greater detail:

\begin{table}[tbh]
\renewcommand{\arraystretch}{1.3}
\vbox{\columnwidth=24pc
\begin{tabular}{|l|l|c|}\hline
\bf Concept 	& \bf Observable	& $~$ \\\hline\hline
creation of dense nuclear matter	& stopping	&  $\surd\quad$ \\\hline
creation of high temperatures		& energy spectra & $\surd\quad$ \\\hline
compression of nuclear matter		& collective flow & $\surd\quad$ \\\hline
hadron source space-time evolution	& particle interferometry & $\surd\quad$ \\\hline
remnants of hadronization		& strangeness enhancement & $\surd\quad$ \\\hline
ashes of plasma				& strangelets and hypermatter & ?$\quad$ \\\hline
restoration of chiral symmetry		& masses/widths of vector mesons & $\surd\quad$ \\\hline
Debye-screening in a QGP		& $J/\Psi$ and $\Psi'$ suppression & $\surd\quad$ \\\hline
radiation of plasma			& direct photons \& thermal dileptons & $\surd\quad$ \\
\end{tabular}
}
\hskip3pc
\caption{\label{tab1} 
Overview of concepts relevant for a QGP search and their 
corresponding observables investigated
at the CERN/SPS. The last column indicates the success of the measurement
and its compatibility with the possible creation of a new form of deconfined
matter.}
\end{table}

\subsection{Multi-strange Baryons}

The relative enhancement of strange and especially multistrange baryons 
with respect to peripheral (or proton induced) 
interactions has been suggested as a signature for the transient 
existence of a QGP-phase \cite{raf8286,koch86,koch88}:  
the main argument being that the (chemical or flavor) equilibration 
times should be much shorter in the plasma phase than in 
a thermally equilibrated hadronic fireball of $T\sim 160\,$MeV.

The dominant production mechanism in an equilibrated (gluon rich) plasma 
phase,  namely the production of $s\overline{s}$ pairs via gluon fusion 
($gg \rightarrow s \overline{s}$) \cite{raf8286},  
should allow for equilibration times similar to the 
interaction time of the colliding nuclei, and to the expected plasma lifetime 
(a few fm/c).

The yields of strange baryons per event calculated in a microscopic
string/hadron transport model (UrQMD) 
are shown in the left frame of figure~\ref{fig_strange_photon} 
as a function of the number 
of participants  for Pb+Pb and p+Pb collisions at 160 GeV/u \cite{sven}.
The $\Lambda+\overline{\Lambda}$- (circles), 
$\Xi^- + \overline{\Xi^-}$- (squares), 
and $\Omega^- + \overline{\Omega^-}$- (triangles) values 
are shown.
The stars correspond to experimental data of the 
WA97 collaboration \cite{strange_data}. Open symbols represent the results 
of the standard UrQMD calculations, whereas 
full symbols exhibit a calculation 
with an enhanced string tension of $\kappa=3\,$GeV/fm, 
for the most central collisions ($N_{\rm part}\ge 300$).
Obviously the standard UrQMD calculation, which can be 
seen as a {\em baseline} of known hadronic physics, strongly underestimates
the (multi-)strange particle yields in central collisions, in particular in
the case of the $\Omega^-$. Only the inclusion of non-hadronic medium effects,
like color-ropes \cite{ropes}, which are simulated by increasing the
string-tension for central collisions, enhances the yield to a level of 
near-compatibility with the data. Similar findings have also been made
in the context of HIJING calculations \cite{hijing}. While these findings
by no means prove the validity of the color-rope approach, they clearly show
the necessity of some kind of medium effect beyond regular binary hadronic
(re)scattering in order to understand the data. Recently, 
hadronic multi-particle interactions in the early dense reaction phase have
been suggested to significantly enhance the yield of anti-protons and 
(anti-)hyperons \cite{rapp,greiner}. It remains to be seen, however, whether
these effects are sufficient to explain the observed $\Omega^-$ enhancement
or other non-hadronic (i.e. deconfinement based) effects need to be 
taken into account.

\subsection{Penetrating Probes: Dileptons and Direct Photons}

\begin{figure}
 \centerline{\epsfxsize=0.5\textwidth\epsfbox{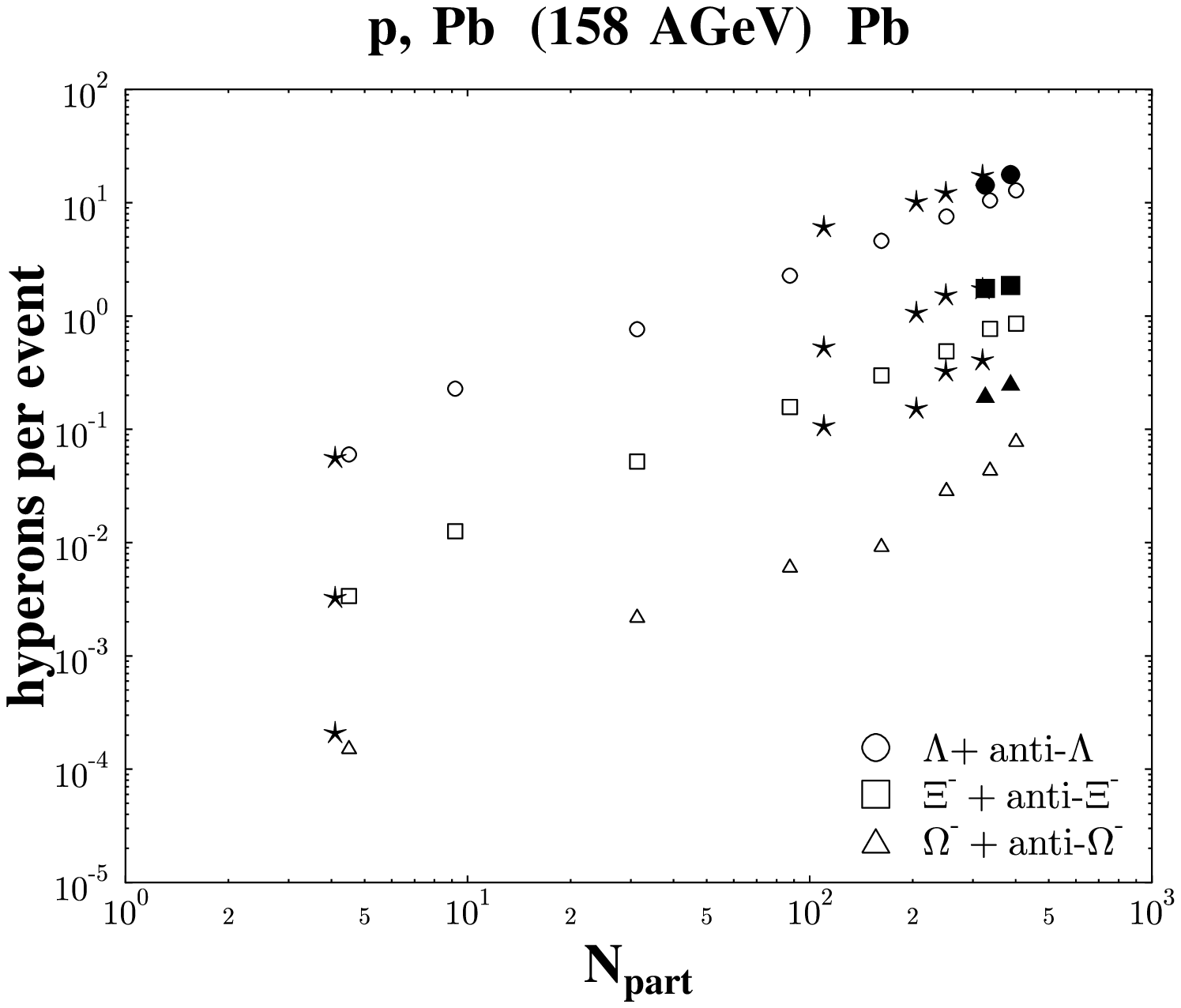} 
\hfill \epsfxsize=0.5\textwidth\epsfbox{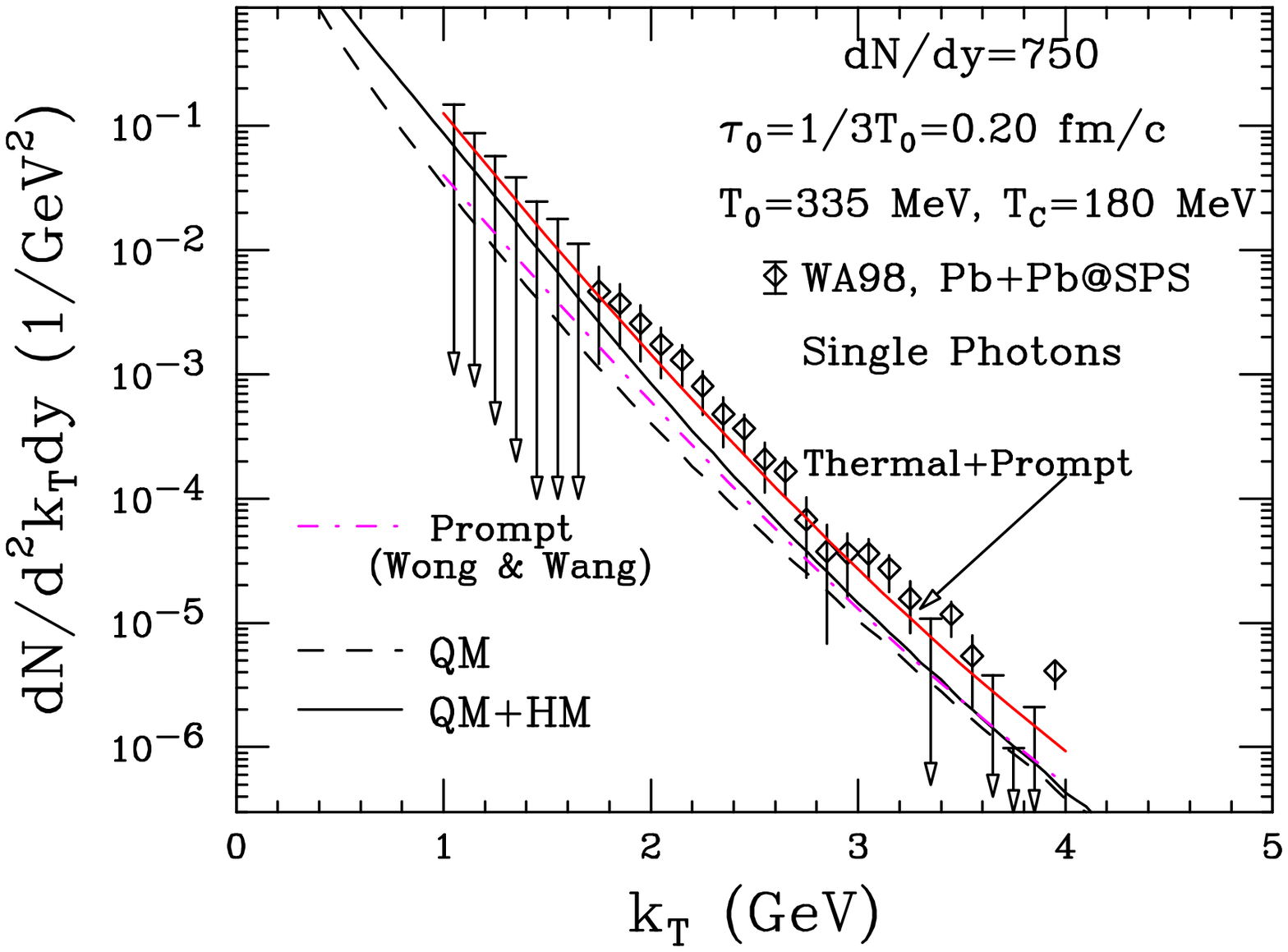}
}
\caption{\label{fig_strange_photon}
Left: excitation function of strange baryon multiplicity
vs. number of participants in a microscopic transport approach. 
Right: Hydrodynamical analysis of direct photon production at the SPS.
}
\end{figure} 

Direct (thermal) photons in a
QGP are created dominantly via
$q\bar{q} \rightarrow \gamma g$ (annihilation) and 
$g q \rightarrow \gamma q$ (Compton scattering).
The production rate and the momentum distribution
of the photons depend on the momentum distributions of quarks, anti-quarks
and gluons in the plasma.
If the plasma is in thermodynamic equilibrium, 
the photons may carry information on this thermodynamic
state at the moment of their production 
\cite{kapusta91a,shuryak78b,sinha83a,hwa85a}.

The main hadronic background processes to compete against are 
pion annihilation $\pi \pi \rightarrow \gamma \rho$ and Compton
scattering $\pi \rho \rightarrow \gamma \pi$ \cite{kapusta91a,song93a}.
The broad $a_1$ resonance may act as an intermediate state in
$\pi \rho$ scattering and thus provide an important 
contribution \cite{xiong92a,song93a} via it's decay into $\gamma \pi$.
In the vicinity of the critical temperature $T_C$ a hadron gas
was shown to
``shine'' as brightly (or even brighter than) a QGP \cite{kapusta91a}.

Hydrodynamical calculations can be used to compare purely hadronic
scenarios of photon radiation
with scenarios involving a first/second order phase transition to a 
QGP. They show a reduction in the temperature of the photon spectrum
in the event of a first order 
phase-transition \cite{alam93a,dumitru95a,neumann95a}.
The right frame of figure~\ref{fig_strange_photon} shows a hydrodynamical
calculation assuming the creation of a QGP and subsequent hadronic
rescattering \cite{dks_photons}. 
Recent estimates of photon production in quark-matter 
(at two loop level) along with the dominant reactions in the hadronic 
matter leading to photons are used. About half of the radiated photons 
are seen to have a thermal origin. The same treatment and the initial 
conditions provide a very good description to hadronic spectra measured 
by several groups, lending additional support to the conclusion that 
quark gluon plasma has been formed in these collisions.  

Dileptons carry
information on the thermodynamic state of the medium at the moment
of production in the very same manner as the direct photons -- 
since the dileptons interact only electromagnetically
they can leave the hot and dense reaction zone basically undistorted, too.
It has been found that at the SPS the invariant mass spectrum of 
low-mass dileptons is 
dominated by pion-pion scattering. Medium modifications to the form-factor
are necessary for a solid understanding of the data -- the current status
of experiment and theory point toward a strong broadening of the $\rho$
resonance without a significant mass-shift toward lower 
masses \cite{sps_dilep}.

\subsection{Chemical Equilibration}

\begin{figure}
 \centerline{\epsfxsize=0.6\textwidth\epsfbox{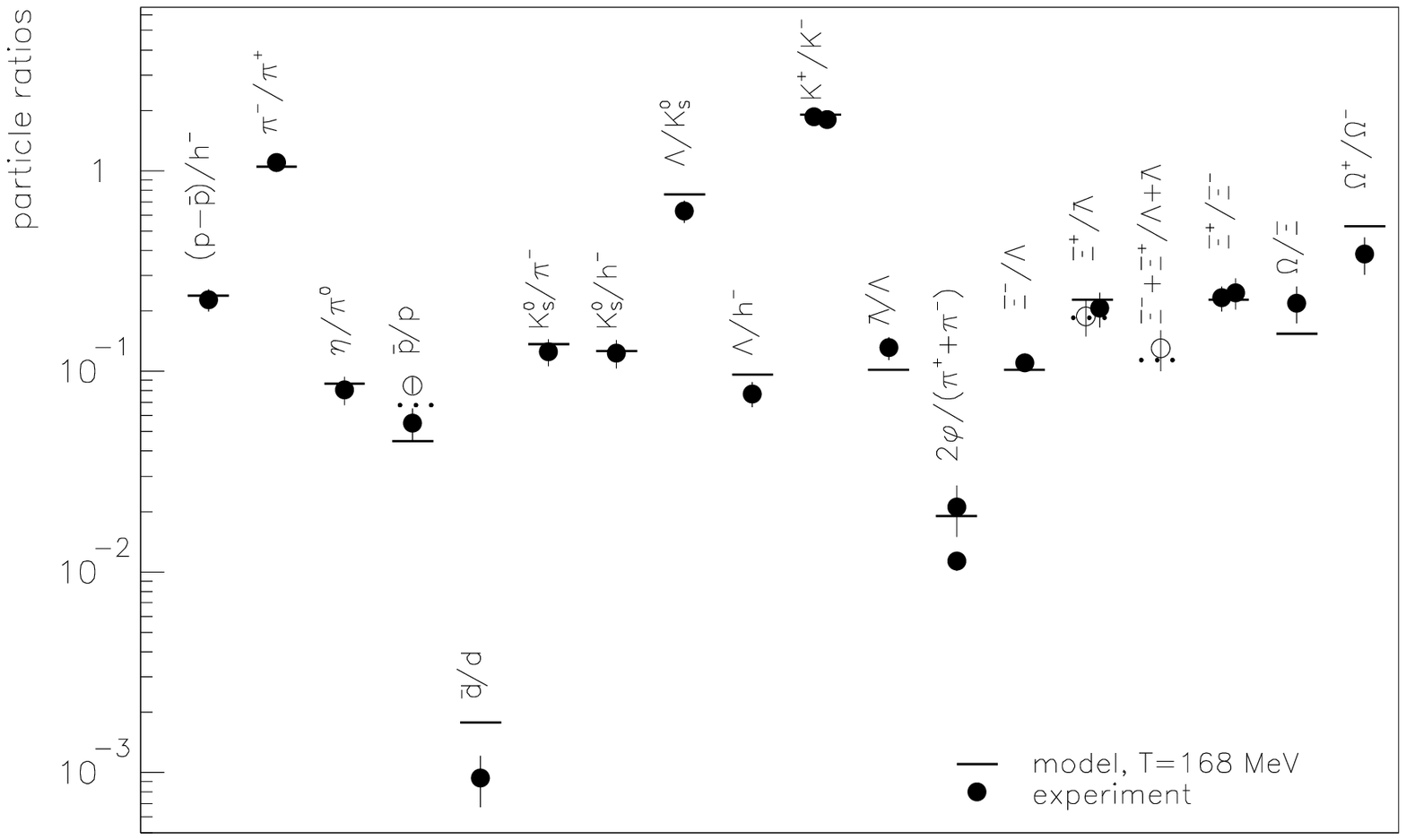} 
\hfill \epsfxsize=0.4\textwidth\epsfbox{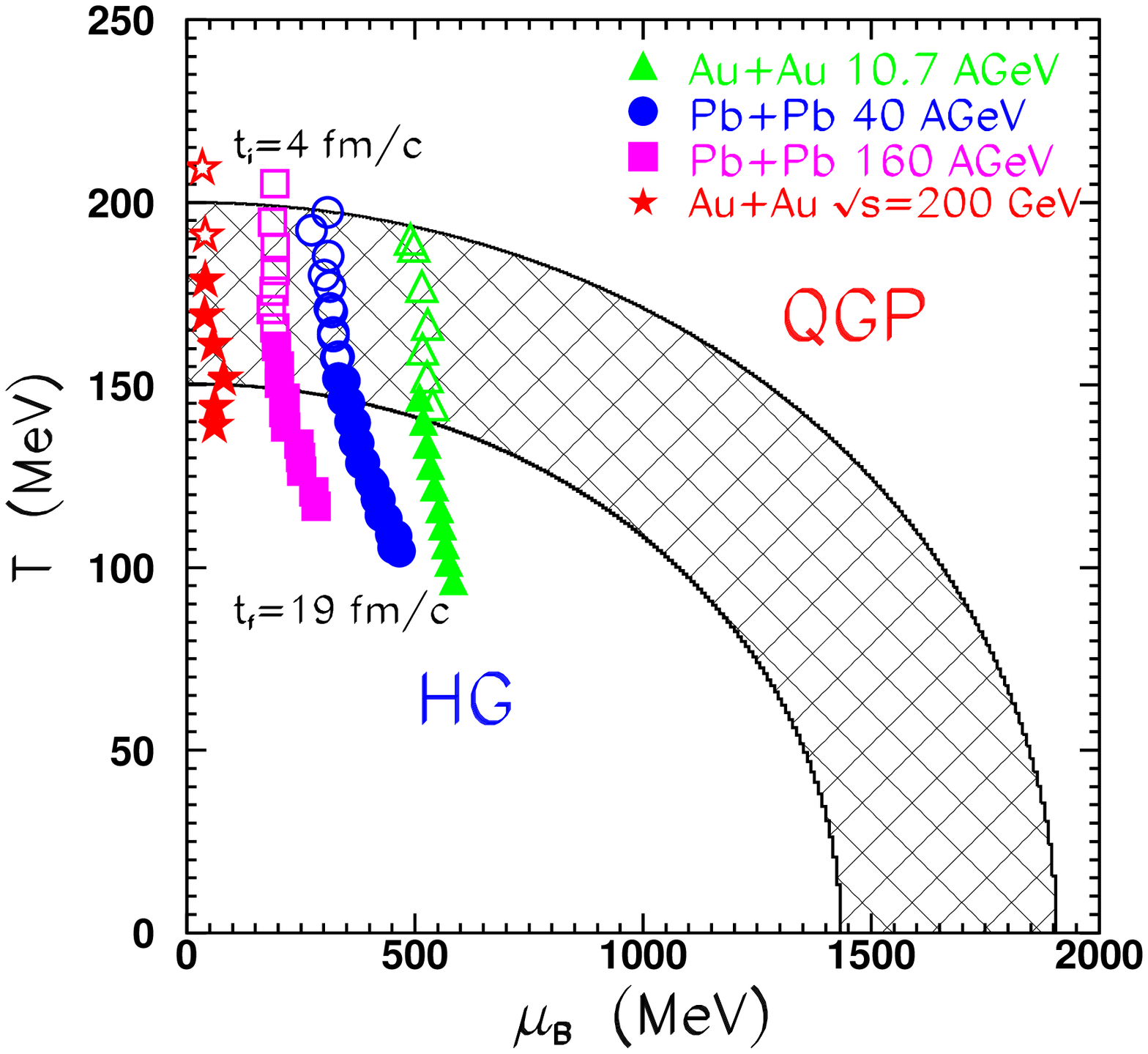}
}
\caption{\label{ratiosfig} Left: statistical model fit to hadron ratios
measured at the SPS. Right: statistical model analysis of the central
cell of a heavy-ion collision calculated with a microscopic transport model.
}
\end{figure}

The success of statistical models in describing the (strange) hadron
abundances and ratios at the CERN/SPS \cite{statmodels}  (see the 
left frame of figure~\ref{ratiosfig}, taken from \cite{heppe})
and the extracted
$\gamma_s$ values close to 1
have led to the common belief that chemical freeze-out in heavy-ion reactions
at the SPS occurred very close to -- or even at hadrochemical equilibrium
and that this state most likely has been created by a hadronizing QGP
\cite{cern_announcement,heinz,braun-munzinger99b}. 
 
However, the estimated chemical equilibration times
may not be sufficiently rapid to cause
chemical equilibration
before hadronization: calculations based on boost-invariant hydrodynamics
with rate-equations for quark production
\cite{kapusta86a,matsui86a,elliott00a},
pQCD rate-equations \cite{biro93a} or
the Parton Cascade Model \cite{geiger93a} all indicate that chemical 
equilibration (and
strangeness saturation) cannot be achieved during realistic life-times
of the deconfined phase. It has been suggested (e.g. in
\cite{kapusta86a,matsui86a}) that the system would be driven toward and
come close to chemical equilibrium in the subsequent hadronic phase
-- a scenario which would help to bridge the gap between the calculations
indicating insufficient equilibration time in the plasma phase and the SPS
data apparently close to chemical equilibrium at chemical freeze-out. 
Recent calculations assuming a hadronizing QGP out of chemical equilibrium
with subsequent hadronic rescatting have shown that rescatting via binary
collisions in the hadronic phase is insufficient to drive the system toward
chemical equilibrium before the expansion of the system leads to chemical
freeze-out \cite{bass_sqm00}. 
However, hadronic multi-particle collisions shortly after
hadronization may have a large impact and drive the system more effectively
toward chemical equilibrium \cite{rapp,greiner} -- a detailed calculation
to verify/falsify this speculation has yet to be performed.

Another question which needs to be raised is whether the temperature
and chemical potential extracted from the
statistical model fits to final state hadron yields or ratios really reflect
the thermodynamic state of the system at one particular time during its 
evolution (i.e. the conditions at chemical freeze-out) or are rather
the result of a superposition of different states, due to individual
hadron species decoupling continuously from the system (as would
be expected from their different mean free paths). The latter 
view is supported by a transport model analysis of the time-evolution
of the temperature and chemical potential in the central
cell of a heavy-ion reaction (see the right frame of figure~\ref{ratiosfig},
taken from \cite{bravina}). However, true progress on this issue can only be
achieved if a model-independent method is found to determine the 
chemical decoupling time of individual hadron species.

\subsection*{SPS: final remarks}
The lessons learned from the SPS experiments may be summarized by the 
following core observations \footnote{a detailed discussion of charmonium
suppression can be found at the end of the RHIC section}:
\begin{itemize}
\item SPS experiments have created a new state of high energy-density
and temperature matter
\item sub-hadronic degrees of freedom offer a viable explanation for many
of the observed phenomena
\item no single observable or measurement is capable of unambiguous proof
for the onset of deconfinement
\item only a combination of observables and experiments will be able to
deliver proof of a new phase of matter
\item the concept of a QGP needs to be rethought: in the SPS
domain (and perhaps even at RHIC) searching for a fireball of quarks
and gluons interacting dominantly 
according to leading order pQCD is unrealistic --
most likely deconfinement will occur and be observed in the form of a fast
evolving phase of partonic degrees of freedom interacting in the domain
of soft non-perturbative QCD.
\end{itemize}

\section{RHIC: first results and expectations}

The Relativistic Heavy Ion Collider has provided an impressive array of
data in its first year of operation and QGP theory faces a stiff challenge 
to keep up with the pace at which new and surprising results emerge.
In the following sections a brief overview on some of the most exciting
and challenging facets of RHIC theory will be given -- a more comprehensive
review on the current status can be found in \cite{mclerran_rhic}.

\subsection{Initial Conditions}

An issue of major importance for understanding the physics at RHIC is how
to describe the initial partonic configuration of the system. In recent
years the {\em Color Glass Condensate} \cite{mlv} has emerged as one of the
leading concepts in that domain: with decreasing $x$ (the fraction of the
longitudinal momentum of the parent hadron), the gluon density in the hadron
wavefunction grows faster than the quark density, giving rise to a 
high-density multiparticle state of gluons. Eventually, the increase
in the number of gluons has to cede and {\em saturation}, 
i.e. a limitation of 
the maximum gluon density per unit phase-space, occurs. Saturation can be 
characterized by an intrinsic momentum scale $Q_s$ (the saturation scale),
below which  non-linear effects in the parton structure function evolution
are expected to slow down and eventually saturate the increase of the gluon
density. The saturation scale $Q_s$ may also provide a natural cut-off
for pQCD based calculations, such as in a parton cascade model.
As we shall see in the following sections, the Color Glass Condensate (CGC)
approach provides a viable explanation for a number of observations made
at RHIC.

\subsection{Hard Processes}

% discuss kharzeev/nardi

\begin{figure}
 \centerline{\epsfxsize=0.5\textwidth\epsfbox{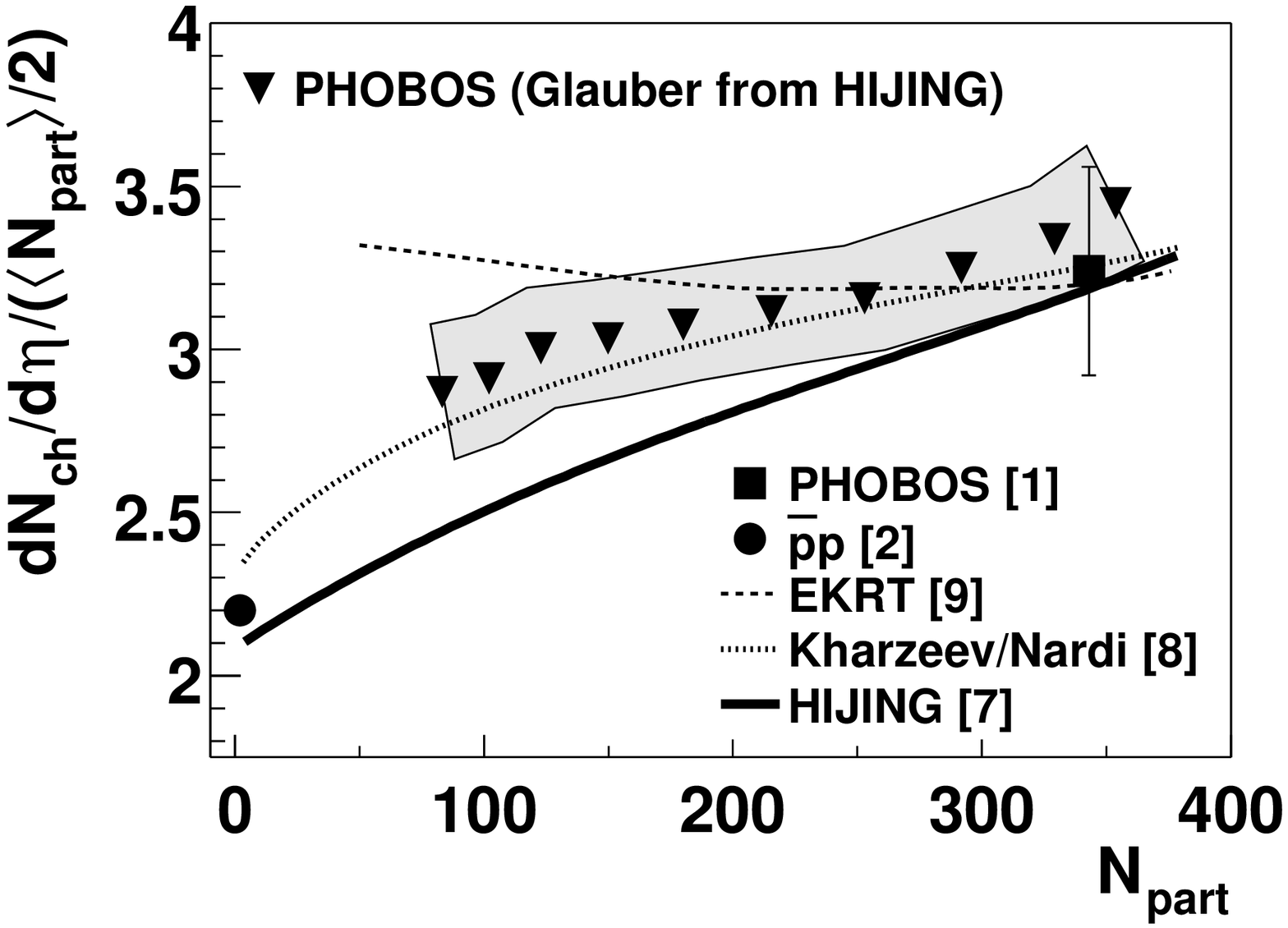} 
\hfill \epsfxsize=0.5\textwidth\epsfbox{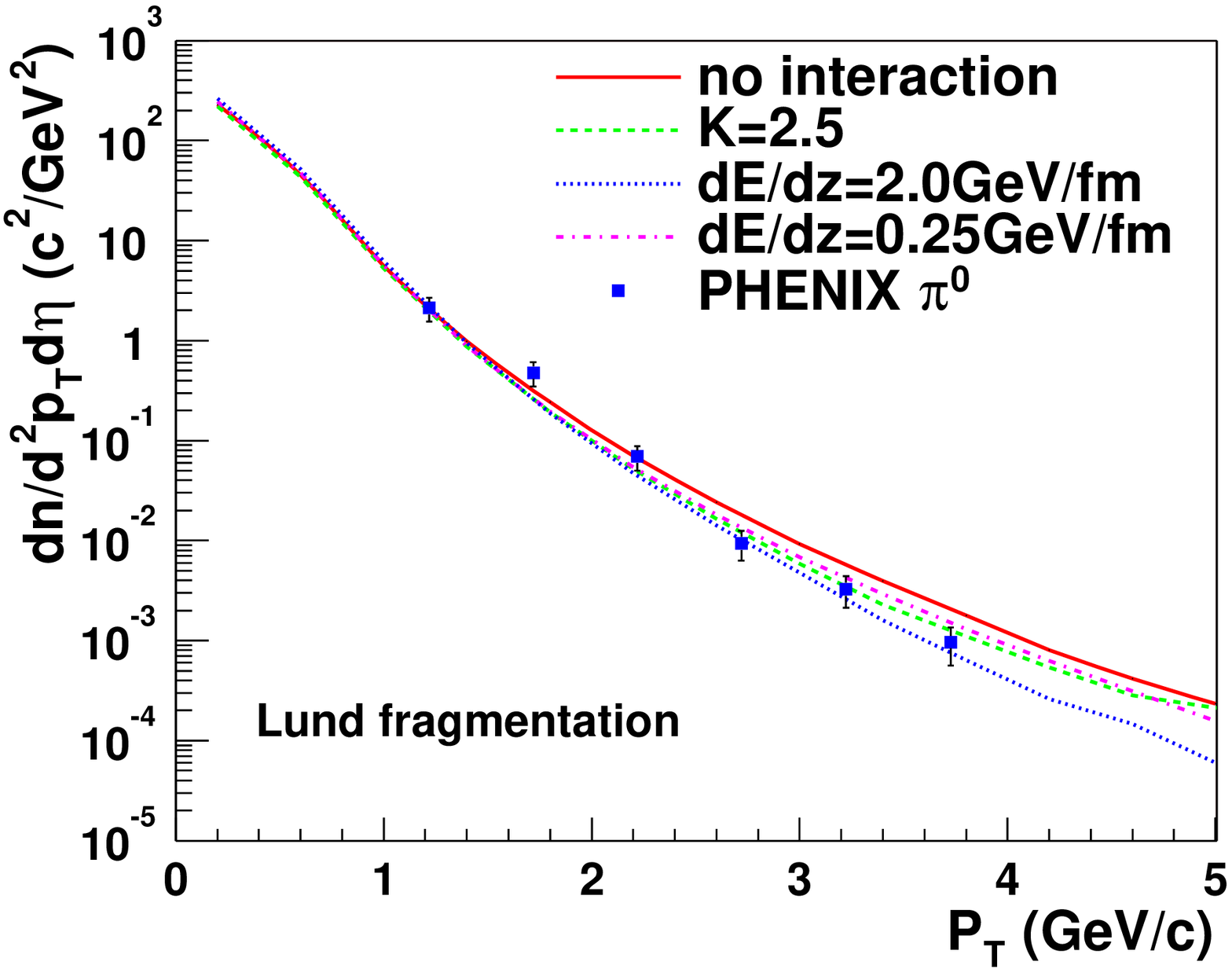}
}
\caption{\label{hard} Left: Charged particle multiplicity per participant
as a function of centrality measured by the PHOBOS experiment compared to
various theoretical calculations. Right: $\pi^0$ transverse momentum spectrum
measured by PHENIX compared to pQCD based calculations employing different
scenarios of partonic energy loss.
}
\end{figure}

One of the first questions to ask concerning the initial RHIC results is 
whether evidence for hard processes can be seen and if so, to what extent
they dominate particle production and the underlying reaction dynamics: 
the PHOBOS data on the charged particle multiplicity at midrapidity and its
centrality dependence \cite{phobos_data} 
has been analyzed in a variety of different
approaches, spanning from saturation model approaches to pQCD mini-jet
production via hard processes \cite{kn,gw,jeon}. 
Of particular note is a simple
analysis carried out in the eikonal approach, decomposing particle production
into a hard component, which scales with the number of collisions
$\langle N_{coll}\rangle$, and a soft component, which scales
with the number of participants $\langle N_{part}\rangle$  \cite{kn,gw}:
\begin{equation}
\frac{dN}{d\eta}\,=\, (1-x)\,n_{pp}\,\frac{\langle N_{part}\rangle}{2}
\, + \, x\, n_{pp}\,\langle N_{coll}\rangle
\end{equation}
with $n_{pp}$ being the charged particle multiplicity in non-single diffractive
$\bar pp$ interactions at the respective $\sqrt{s}$.
Using parameters extracted from a Glauber calculation and  
fitting the parameter $x$ to the PHOBOS data 
this analysis yields a fraction of 37\%
of the produced particles stemming from hard processes at $\sqrt{s}=130~$GeV. 
The centrality dependence of the charged particle multiplicity is 
nicely reproduced as well - see the left frame figure~\ref{hard}
(taken from \cite{phobos_data}).
However,
an analogous analysis utilizing the saturation approach of the Color Glass
Condensate yields similar agreement with the data, when employing a 
centrality dependent saturation scale $Q_s(b)$
\cite{kn}:
\begin{equation}
\frac{2}{N_{part}}\,\frac{dN}{d\eta}\,\simeq \, 0.82  \ln\left( 
\frac{Q^2_s(b)}{\Lambda^2_{QCD}} \right)
\end{equation}

Since both, the conventional QCD and the high-density QCD (i.e.
saturation) approach are able the describe the data, it is currently difficult 
to distinguish between these two approaches. Comparisons to other data,
e.g. transverse momentum spectra, will be able to resolve this ambiguity.

%discuss jet quenching, newer developments

Partons created in a hard scattering can act as very efficient probes
of a QGP: a parton traversing a color confined medium of hadrons sees 
a relatively transparent system. However, a parton passing through a hot
medium of deconfined degrees of freedom is expected to loose a large
amount of energy via gluon radiation. Since the radiated gluons may couple
to each other, the energy loss is predicted to be proportional to the square
of the length traversed in the deconfined medium \cite{quench} -- this
phenomenon is commonly being referred to as {\em jet-quenching}.
RHIC experiments have established a significant suppression
of high-$p_T$ hadrons produced in central A+A collisions compared to those
produced in peripheral A+A or binary scaled p+p reactions, indicating
a strong nuclear medium effect, compatible with the jet-quenching 
predictions\cite{phenix,star}.
Early calculations predicting a fairly large rate of energy loss
around 2~GeV/fm have been shown to be incompatible with the measured 
spectra at RHIC \cite{phenix,star}, whereas
newer calculations based on the GLV formalism \cite{glv1} for ``thin'' plasmas
and including finite kinematic effects provide a far better agreement
with the data \cite{glv2} -- similar to the assumption of an rather
small energy loss around 0.25~GeV/fm.

However, alternative scenarios for partonic energy loss have to be considered:
the parton may change its momentum through hard scattering in the deconfined
medium \cite{nara} or after hadronizing through soft hadronic rescattering 
\cite{bass_turin}. Figure~\ref{hard} (taken from \cite{nara})
shows a parton cascade calculation which
compares standard parton energy loss calculations with a collisional 
energy loss scenario - both approaches are compatible with 
the PHENIX data \cite{phenix}.

\subsection{Balance Functions and Fluctuations}

While there is no doubt about the importance of partonic degrees of
freedom for the initial, early, reaction stages, one of the crucial questions
is how long the deconfined state may have actually existed and whether this
time-span is long enough for thermalization and collective effects
to occur. Balance functions offer a unique
model-independent formalism to probe the time-scales of a deconfined phase
and subsequent hadronization \cite{balance}.
If a long-lived QGP has been formed, a large number of quark-antiquark pairs
need to be created close to the hadronization time, mainly due to
entropy conservation constraints.
This late-stage production of quarks could be attributed to three mechanisms:
formation of hadrons from gluons, conversion of the non-perturbative vacuum
energy into particles, or hadronization of a quark gas at constant
temperature. Hadronization of a quark gas should approximately conserve the net
number of particles due to the constraint of entropy conservation. Since
hadrons are formed of two or more quarks, creation of quark-antiquark pairs
should accompany hadronization. All three mechanisms for late-stage quark
production involve a change in the degrees of freedom. Therefore, any signal
that pinpoints the time where quarks first appear in a collision would provide
valuable insight into understanding whether a novel state of matter has been
formed and persisted for a substantial time.

\begin{figure}
 \centerline{\epsfxsize=0.5\textwidth\epsfbox{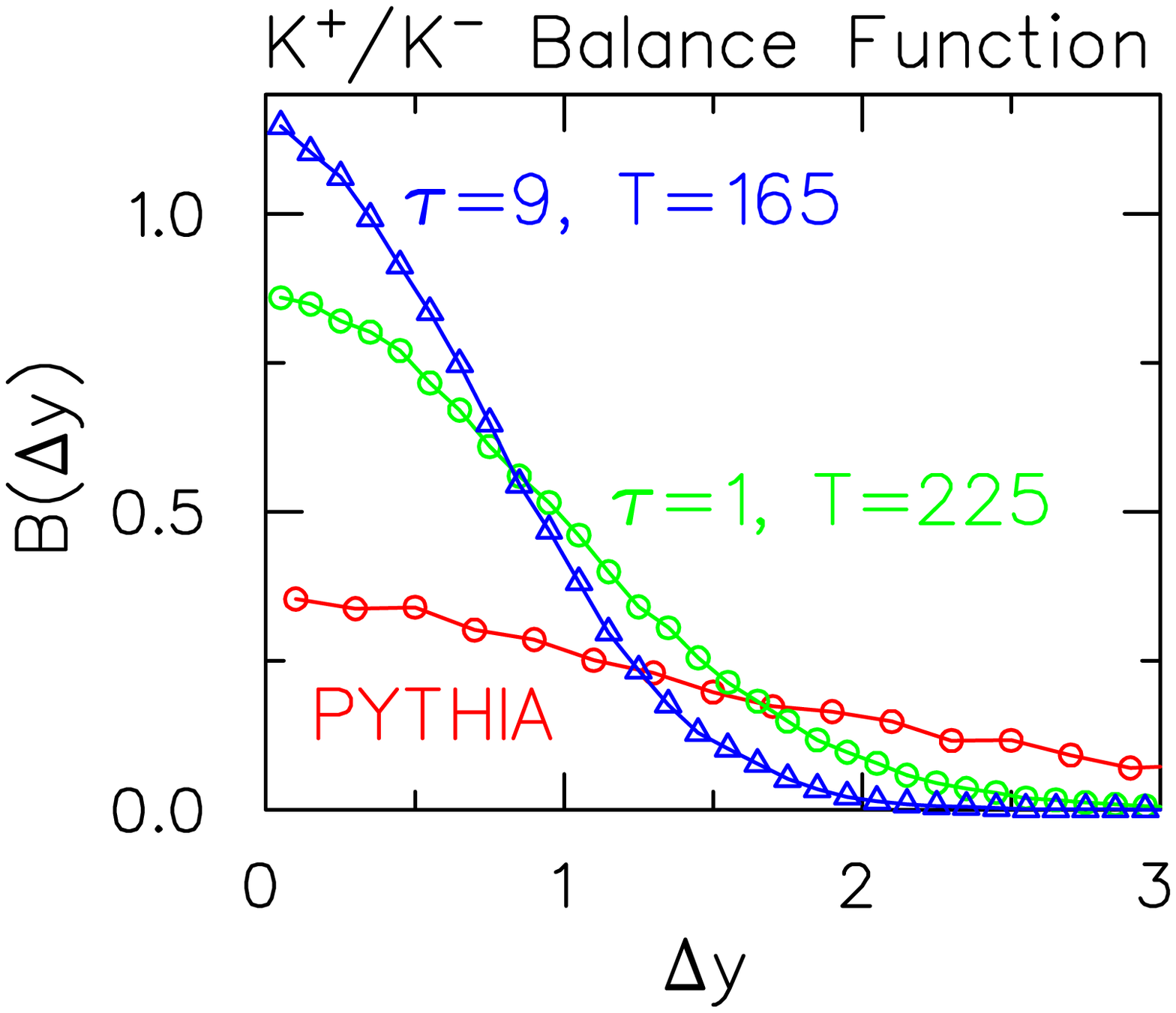} 
\hfill \epsfxsize=0.5\textwidth\epsfbox{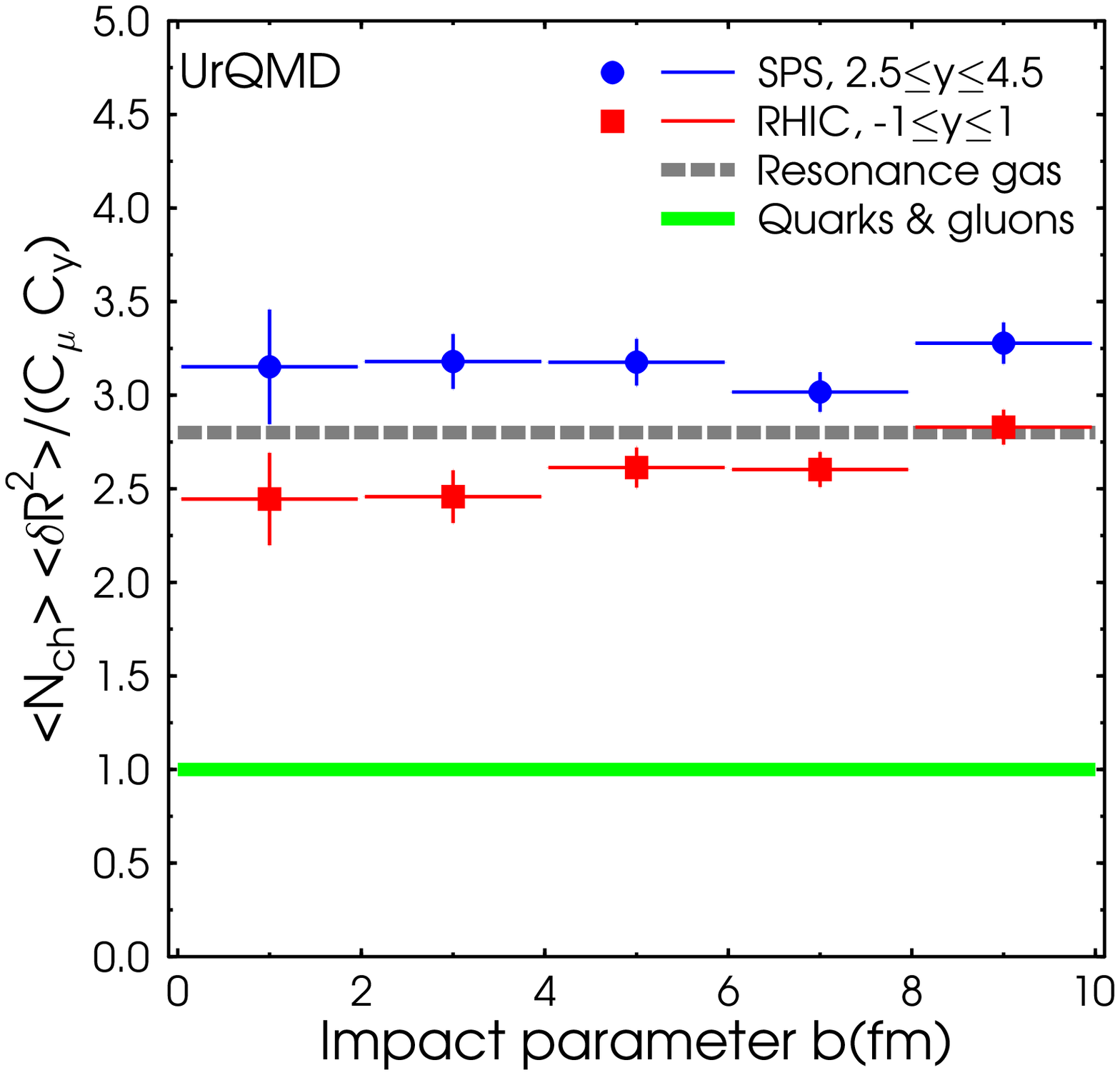}
}
\caption{\label{balfig} Left: $K^+/K^-$ balance functions calculated for
different hadronization times and scenarios. Right: Transport model
calculation of charged particle fluctuations compared to analytical
predictions for a resonance gas and a QGP.
}
\end{figure} 

The link between balance functions and the time at which quarks are created has
a simple physical explanation. Charge-anticharge pairs are created at the same
location in space-time, and are correlated in rapidity due to the strong
collective expansion inherent to a relativistic heavy ion collision. Pairs
created earlier can separate further in rapidity due to the higher initial
temperature and due to the diffusive interactions with other particles. The
balance function, which describes the momentum of the accompanying
antiparticle, quantifies this correlation. 
The balance function describes the conditional
probability that a particle in the bin $p_1$ will be accompanied by a particle
of opposite charge in the bin $p_2$:
%\vspace{-3mm}
\begin{equation}
\label{balancedef_eq}
B(p_2|p_1) \equiv \frac{1}{2}\left\{
\rho(b,p_2|a,p_1)-\rho(b,p_2|b,p_1) +\rho(a,p_2|b,p_1)-\rho(a,p_2|a,p_1)
\right\},
\end{equation}
%\vspace{-3mm}
where $\rho(b,p_2|a,p_1)$ is the conditional probability of observing a
particle of type $b$ in bin $p_2$ given the existence of a particle of type $a$
in bin $p_1$.  The label $a$ might refer to all negative kaons with $b$
referring to all positive kaons, or $a$ might refer to all hadrons with a
strange quark while $b$ refers to all hadrons with an antistrange quark. 

The left frame of figure~\ref{balfig} shows $K^+/K^-$ balance functions 
as predicted in a simple
Bjorken thermal model for two hadronization temperatures, 
225 MeV and 165 MeV as well as for fragmenting strings (utilizing PYTHIA), 
which would represent the hadronization scenario of a hadronic/string picture
similar to that of RQMD or UrQMD.
Since particles from cooler systems have smaller thermal velocities,
they are more strongly correlated in rapidity and result in narrower balance
functions. A strong sensitivity to the hadronization temperature 
and time can be clearly observed.
The STAR collaboration has recently published first results on a 
charged particle balance function analysis at RHIC \cite{gary}: 
a striking centrality dependence is observed, with the width of the
balance function decreasing as a function of increasing centrality, in line
with what to expect in the case of a long-lived deconfined phase.

The size of the average fluctuations of net baryon number and electric charge 
in a finite volume of hadronic matter differs widely between the confined 
and deconfined phases, due to the different value of the elementary charge
carried by quarks vs. hadrons. 
These differences may be exploited as indicators of 
the formation of a quark-gluon plasma in relativistic heavy-ion collisions, 
{\em if} the fluctuations created in the initial state survive 
until freeze-out \cite{asakawa,jeonkoch}. 
The right frame of figure~\ref{balfig} (taken from \cite{blfluc}) 
compares transport 
calculations in the framework of a string/hadron model 
for SPS and RHIC with analytical predictions for charge 
fluctuations in a hadronic resonance gas and a parton gas \cite{blfluc}. 
Since the 
transport model does not contain any deconfined degrees of freedom,
the numerical calculation agrees well with the resonance gas prediction.

Interestingly, sofar all experimental analysis for SPS and RHIC agree 
with the hadron gas prediction, giving no indication at all about a possible 
deconfined phase. Since many other observables are compatible with the
assumption of deconfinement, it may very well be  that the dynamics of
hadronization strongly affects the charge 
fluctuation observables and masks the deconfined phase.

Under certain assumptions, charge fluctuations can be directly 
expressed in terms of balance functions \cite{pratt_jeon}:
\begin{eqnarray}
\frac{\langle(Q-\langle Q\rangle)^2\rangle}{\langle N_{\rm ch}\rangle}
=1-\int_0^Y d\Delta y~B(\Delta y | Y)
+O\left(\frac{\langle Q\rangle}{\langle N_{\rm ch}\rangle}\right)
\label{eq:relation}
\end{eqnarray}
The size of the correction term is usually less than 5\,\% for electrical
charge fluctuations, since 
the number of produced charges is much larger than the net
charge. However, for baryon number fluctuations 
the additional term is not negligible, even at RHIC.

Comparing the apparent lack of QGP indications of the charge fluctuation 
observables with the promising results of the balance function analysis
one is lead to speculate whether relevant dynamical information of the 
two particle correlation is averaged out by integrating over the correlation
function in order to obtain the fluctuation observable.

\subsection{Collective Flow}

\begin{figure}
 \centerline{\epsfxsize=0.5\textwidth\epsfbox{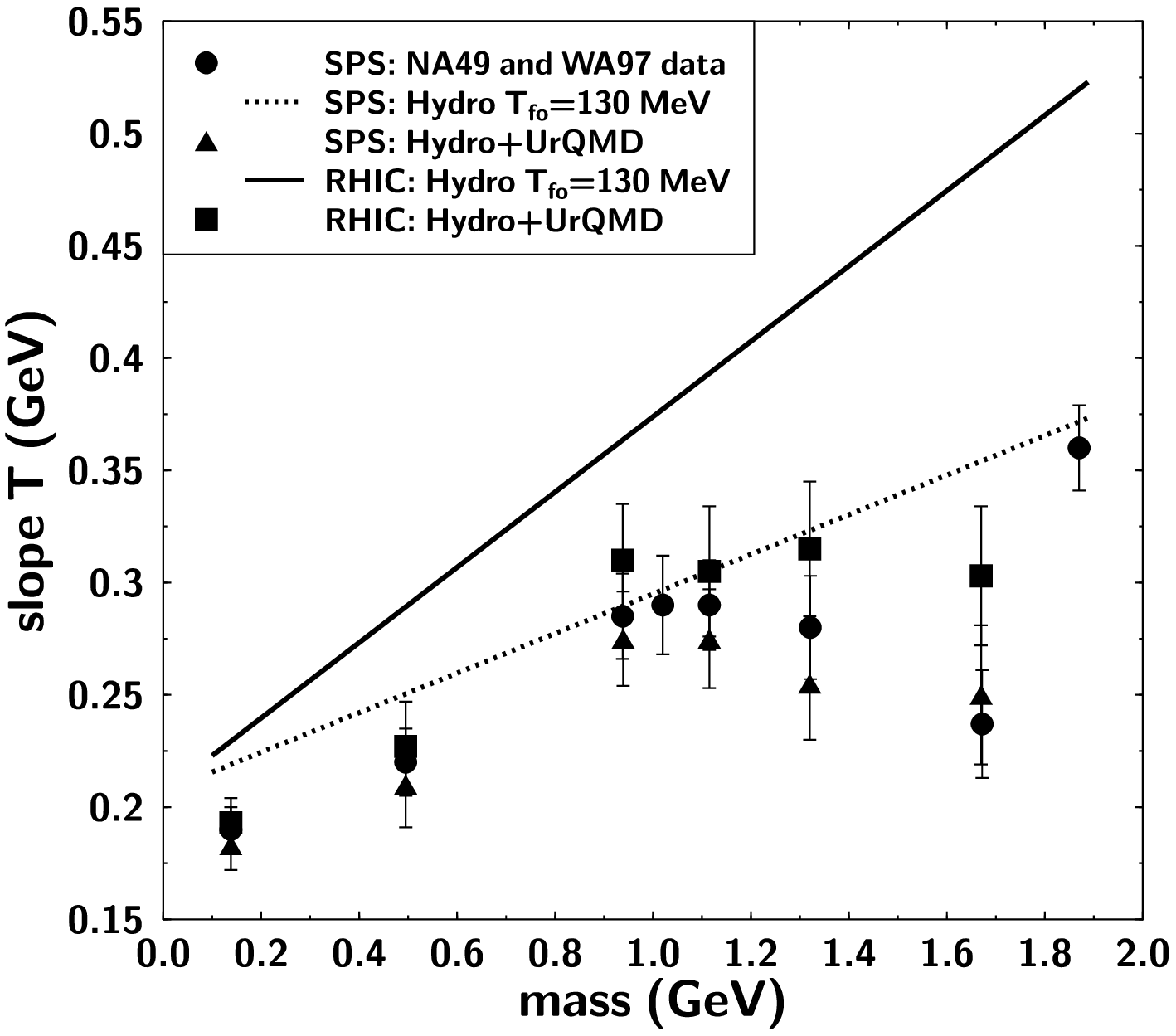} 
\hfill \epsfxsize=0.5\textwidth\epsfbox{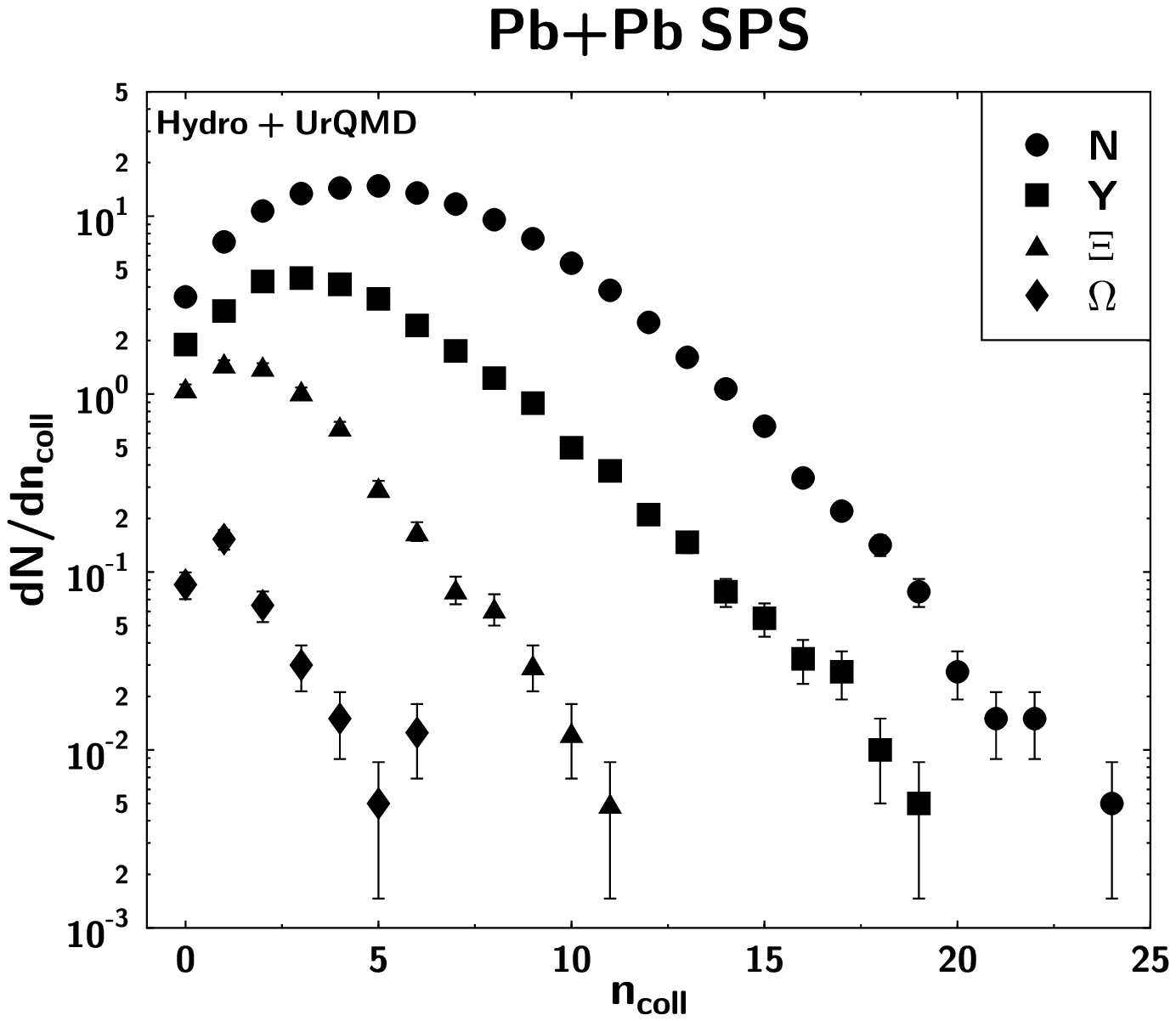}
}
\caption{\label{mslope} Left: Inverse hadron $m_T$ slopes at $y_{c.m.}=0$.
The lines depict pure hydrodynamics whereas the symbols
refer to data and hydro+micro calculations. Right: collision number
distributions for different baryon species in the hadronic phase of a 
hydro+micro calculation.
}
\end{figure} 

The study of deconfinement and a subsequent phase-transition to 
deconfined hadronic matter poses a great challenge to microscopic transport
models. In most approaches hadronization is a uni-directional process
and the equation of state of the system ill-defined. One possible remedy 
is to use a hydrodynamical approach for the early deconfined phase of the 
reaction and subsequent phase-transition, coupled with a microscopic 
calculation for the later, hadronic, reaction phase in which the hydrodynamical
assumptions are not valid any longer \cite{hu_main}. In the following,
such a combined macro+micro model will be used to study elliptic flow
as well as the flavor- and 
mass-dependence of hadronic
slope parameters (ie. radial flow):

Radial and elliptic flow in non-central heavy ion collisions can constrain 
the effective Equation of State(EoS) of the excited nuclear matter.
It has been shown that for an EoS with a first order phase transition, 
the above mentioned hybrid macro+micro models reproduce 
both the radial and elliptic flow data at the SPS \cite{hu_main,teaney}.
The centrality dependence
of the elliptic flow coefficient $v_2$ exhibits a strong sensitivity to
the underlying EoS, which may help to constrain the EoS  \cite{teaney2}. 
In addition, a 
number of features of the RHIC data can also be explained within the hybrid 
approach \cite{teaney2}:
the observed elliptic flow and its dependence on $p_{T}$ 
and mass, the anomalous $\bar{p}/\pi^{-}$ ratio for 
$p_{T} \approx 2.0$ GeV, and the difference in the impact parameter 
dependence of the $\bar{p}$ and $\phi$ slope parameters. 
For an EoS without the hard and soft features of the QCD phase 
transition, the broad consistency with the data is lost. 

The left frame of figure~\ref{mslope}
displays the inverse slope parameters $T$ obtained by an
exponential fit to
$dN_i/d^2m_Tdy$ in the range $m_T-m_i<1$~GeV for SPS and RHIC in a hybrid
hydro+micro calculation and
compares them to SPS data \cite{strange_data}.
The trend of the data, namely the ``softer'' spectra of $\Xi$'s and
$\Omega$'s as compared to a linear $T(m)$ relation
is reproduced
reasonably well. This is in contrast to ``pure'' hydrodynamics with kinetic
freeze-out on a common hypersurface (e.g.\ the $T=130$~MeV isotherm), where
the stiffness of the spectra increases linearly with mass
as denoted by the lines in
fig.~\ref{mslope}.
When going from SPS to RHIC energy, the model discussed here generally
yields only a slight increase of the inverse slopes,
although the specific entropy is larger by a factor of 4-5~!
The reason for this behavior is the first-order phase transition that
softens the transverse expansion considerably.   

The softening of the spectra is caused by
the hadron gas emerging from the hadronization of the QGP
being almost ``transparent'' for the multiple
strange baryons.
This can be seen by calculating the average number of collisions
different hadron species suffer in the hadronic phase (see right
frame of fig.~\ref{mslope}):
whereas $\Omega$'s suffer on average only one hadronic interaction,
$N$'s and $\Lambda$'s suffer approx. 5--6
collisions with other hadrons before they freeze-out.

Thus, one may conclude that
the spectra of $\Xi$'s and especially $\Omega$'s  are practically
unaffected by the hadronic reaction stage and closely resemble
those on the phase
boundary. They therefore act as probes of the QGP expansion prior
to hadronization and can be used to measure the expansion rate
of the deconfined phase.

\subsection{Two Particle Interferometry}

\begin{figure}
 \centerline{\epsfxsize=0.45\textwidth\epsfbox{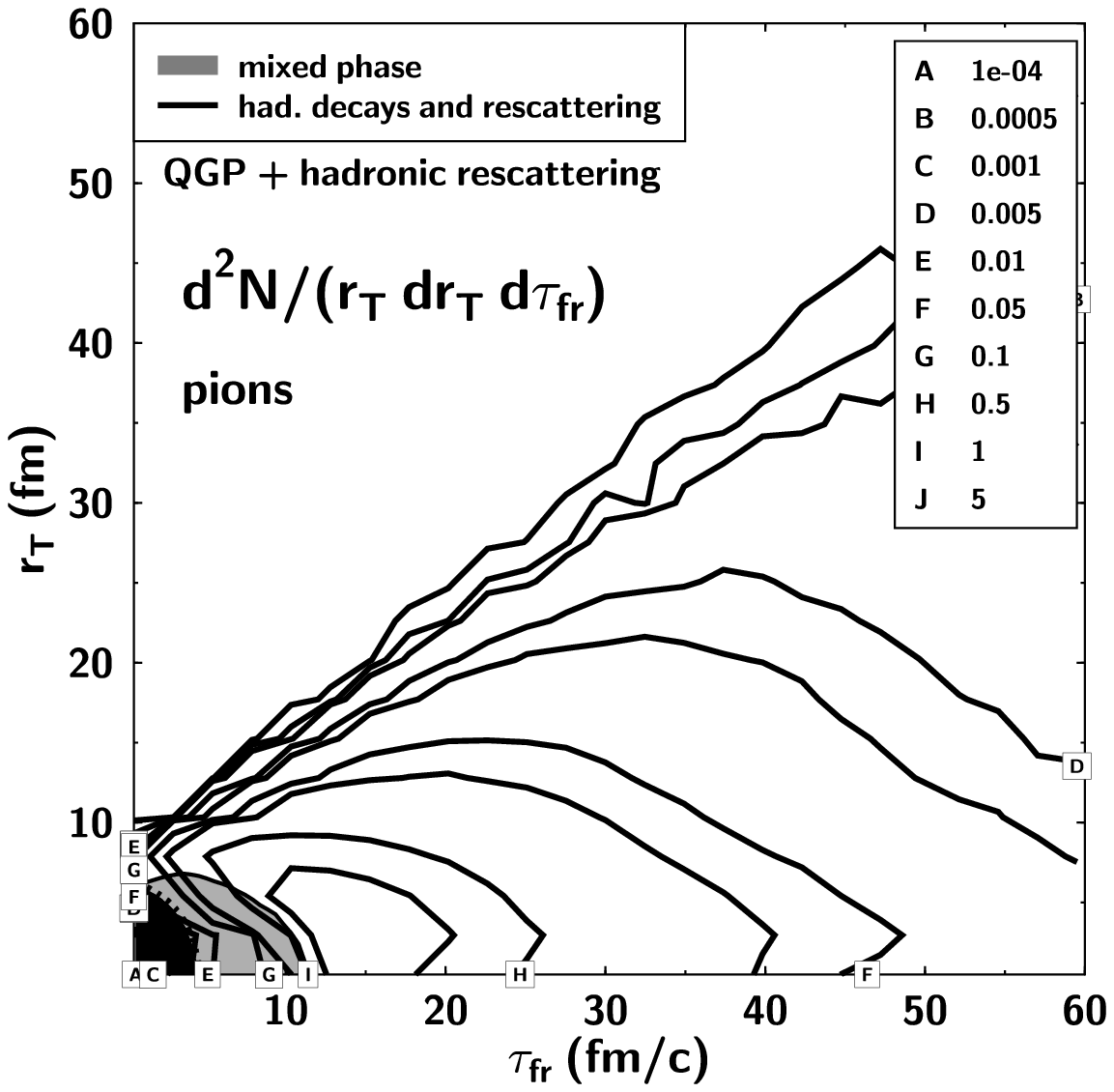}
\hfill \epsfxsize=0.55\textwidth\epsfbox{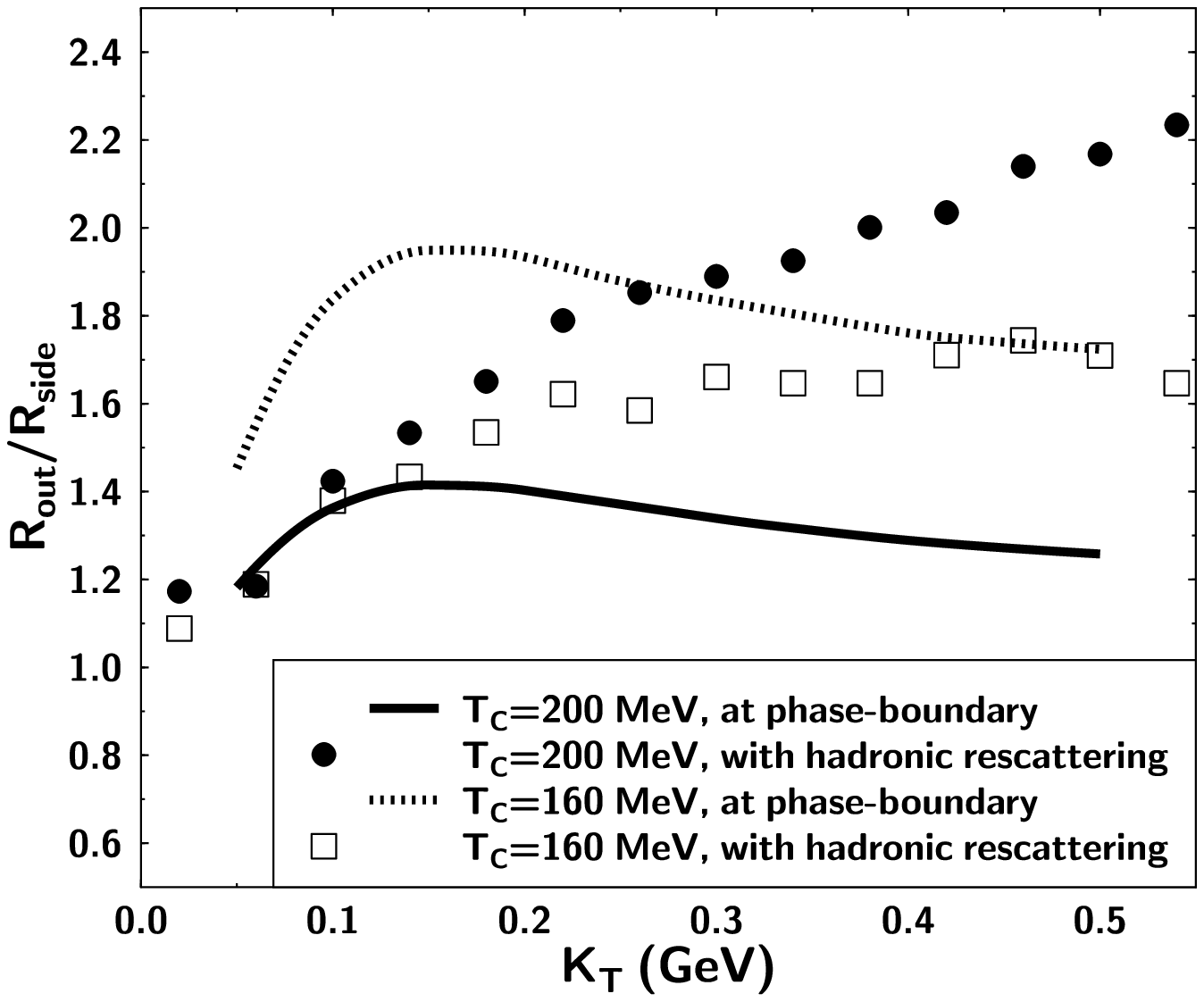} 
}
\caption{\label{rttf} Left: Freeze-out transverse radius and time distribution
$d^2N/(r_Tdr_Tdt_{fr})$ for pions at RHIC. 
Right: $R_{\rm out}/R_{\rm side}$ for RHIC initial conditions,
as a function of $K_T$ at freeze-out
(symbols) and at hadronization (lines).
}
\end{figure} 

Kinetic freeze-out, at which all (elastic) interactions cease, terminates 
the evolution of the reaction discussed in figure~\ref{ttover}. However,
the freeze-out time of the system is ill-defined, since interactions
cease locally on a particle per particle basis.
The left frame of fig.~\ref{rttf} 
shows the distribution of freeze-out points of pions
in the forward light-cone, as well
as the hadronization hypersurface from where all hadrons emerge in a 
hydro+micro approach.
Freeze-out is seen to occur in a {\em four-dimensional} region within the
forward light-cone rather than on a three-dimensional ``hypersurface'' in
space-time. Similar results have
also been obtained within other microscopic transport models~\cite{microFO}
when the initial state was not a QGP.
It is clear that the hadronic system disintegrates  slowly (as
compared to e.g.\ the hadronization time), rather than emitting a ``flash''
of hadrons (predominantly pions) in an instantaneous decay.

A first order phase transition
leads to a prolonged hadronization time as compared to a cross-over
or ideal hadron gas with no phase transition, and has been related to unusually
large Hanbury-Brown--Twiss (HBT) radii~\cite{pratt86,schlei,dirk1}.
The phase of coexisting hadrons and QGP reduces the ``explosivity'' of the
high-density matter before hadronization, extending the emission duration of
pions~\cite{pratt86,schlei,dirk1}.
This phenomenon should then depend on the
hadronization (critical) temperature $T_c$ and the latent heat of the
transition. For recent reviews on this topic we refer
to~\cite{reviews,wiedemannrep}.

It has been suggested that the ratio $R_{\rm out}/
R_{\rm side}$ should increase strongly once the initial
entropy density $s_i$ becomes substantially larger than that of the hadronic
gas at $T_c$~\cite{dirk1}.
The strong $T_C$ dependence
of $R_{\rm out}/R_{\rm side}$ in such a purely hydrodynamical scenario
can be seen in the right frame of  fig.~\ref{rttf} (solid and dotted lines). 
Here, the $R_{\rm out}/R_{\rm side}$ ratio at hadronization (calculated
in a hydrodynamical scenario) is compared 
to the ratio after subsequent hadronic rescattering and freeze-out 
in a combined hydro+micro approach \cite{hbt_prl}. 
Clearly, up to $K_T\sim200\,$MeV (with $K_T=(p_{1,T}+p_{2,T})/2$)
$R_{\rm out}/R_{\rm side}$ is independent of $T_c$, 
if hadronic rescatterings are taken into account.
Moreover, at higher $K_T$ the dependence on $T_c$ is even reversed:
for high $T_c$ the $R_{\rm out}/R_{\rm side}$ ratio even exceeds that
for low $T_c$.
A higher $T_c$ speeds up hadronization but on the other hand prolongs
the dissipative hadronic phase that dominates the HBT radii. This is because
during the non-ideal hadronic expansion the scale of spatial homogeneity
of the pion density distribution increases, as the pions fly away from the
center, but the transverse flow can hardly increase to counteract.
Therefore, after hadronic rescattering
$R_{\rm out}/R_{\rm side}$ does not drop towards
higher $K_T$ (in the range $K_T\lton 3m_\pi$).

For central collisions of Au nuclei at
$\sqrt{s}=130A$~GeV, data from STAR gives
$R_{\rm out}/R_{\rm side}\simeq1.1$ at small $K_T$~\cite{Lisa01}.
With increasing $K_T$ the data remain flat or even decrease slightly -- a 
trend which is currently not understood theoretically and cannot be
reproduced by any of the dynamical models on the market.
Kaon interferometry may provide crucial insight into this so-called
{\em HBT-puzzle} at RHIC \cite{kaon_hbt}: 
the kaon phase-space density is much lower than
that of the pions, thus dramatically reducing higher order correlation 
effects. In addition, high-$p_t$ kaons are emitted to a larger fraction 
directly from the phase-boundary, increasing the sensitivity to the
QGP equation of state. 

\subsection{Dileptons and Charm}

\begin{figure}
 \centerline{\epsfxsize=0.5\textwidth \epsfbox{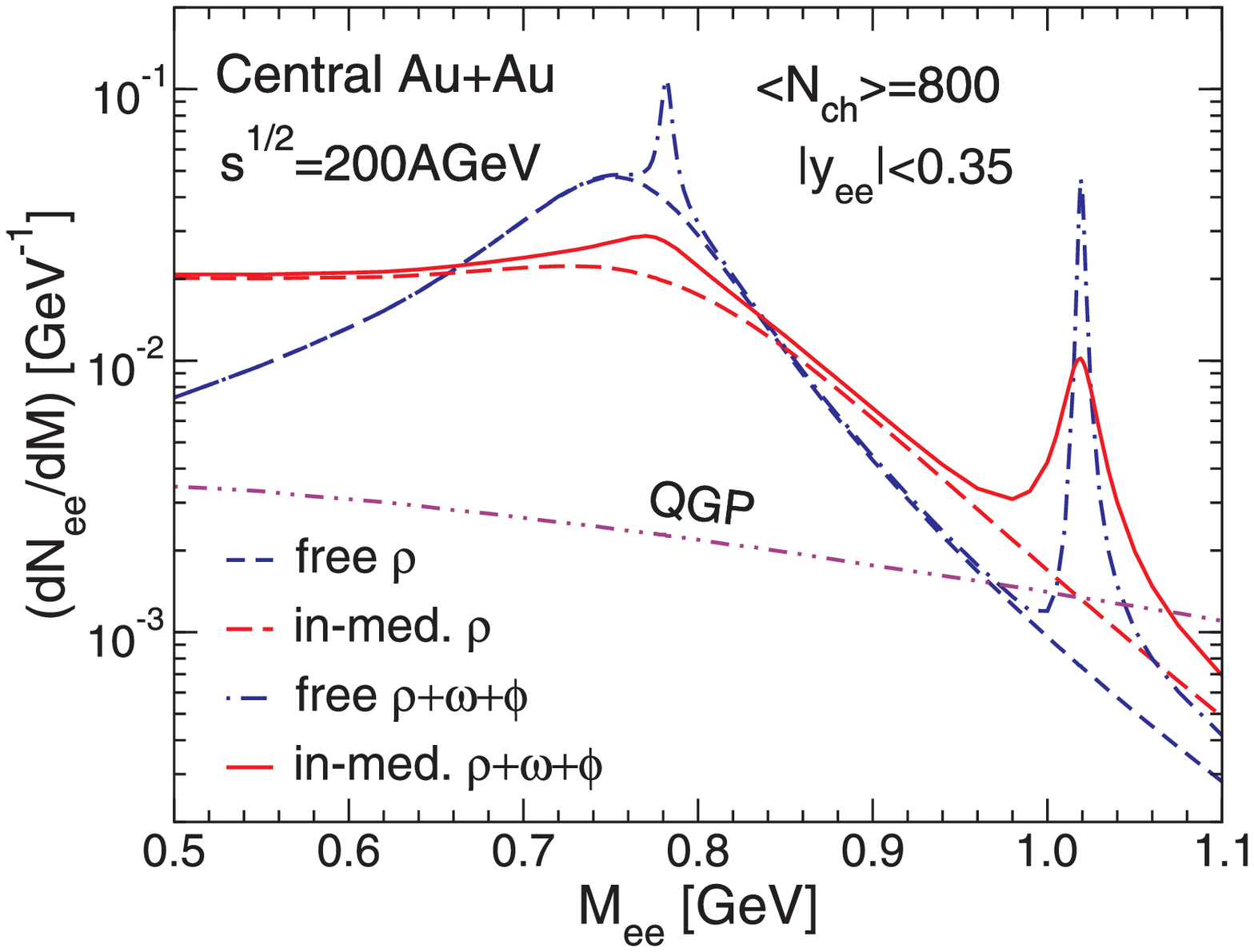} 
\hfill \epsfxsize=0.5\textwidth\epsfbox{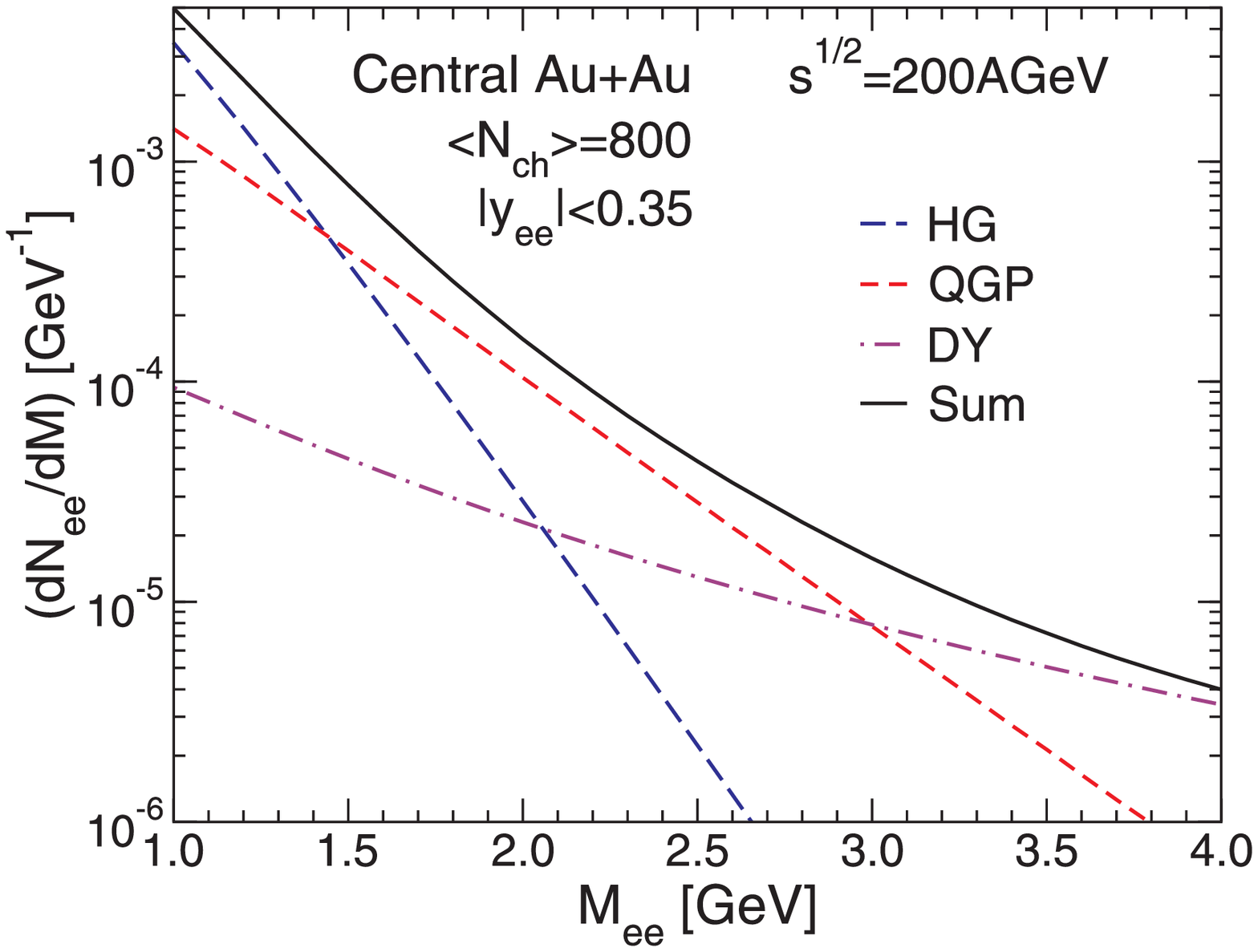}
}
\caption{\label{rapp_rhicfig} Left: Invariant mass spectrum of low mass
dileptons calculated with and without hadronic medium modifications, 
including possible contributions from thermal dileptons stemming from
a QGP. Right: Mass spectrum of intermediate mass dileptons -- around
$M_{ee}\simeq 2.5$~GeV the contribution of QGP radiation dominates above
the hadronic background and Drell-Yan processes.  
}
\end{figure} 

First dilepton and charm measurements will soon be forthcoming at RHIC
and are highly anticipated:
in the dilepton sector, current theoretical expectations are that in the
low mass region thermal dileptons from a QGP 
will be strongly suppressed compared
to hadronic contributions -- mainly from in-medium pion-pion annihilation.
Current state of the art calculations point towards a strong broadening
of vector mesons in medium \cite{rapp_rhic} 
-- the $\rho$ and $\omega$ peaks should be almost
completely dissolved and the $\phi$ significantly broadened (see left
frame of figure~\ref{rapp_rhicfig}, taken from \cite{rapp_rhic}).
At intermediate invariant masses (around 2 -- 2.5~GeV) thermal dileptons
radiated from a QGP start to outshine hadronic sources 
\cite{ruuskanen91a,ruuskanen92a,rapp_rhic}, while 
for higher masses the yield of Drell--Yan processes
exceeds that of thermal dileptons from a QGP (see right frame 
of figure~\ref{rapp_rhicfig}). Open charm may give a large contribution
to the spectrum in the intermediate mass range as well -- however independent
measurements of open charm will allow to subtract that contribution from
the invariant mass spectrum and thus will allow a measurement of the thermal
contribution \cite{gavin96d,rapp_rhic}.

In the charmonium sector, predictions on what to expect at RHIC vary
widely -- from total suppression to a strong enhancement of the $J/\psi$ and
$\psi'$ yield! Originally it was predicted
that the suppression of
heavy quarkonia-mesons could provide one of the signatures for
deconfinement in QCD at high temperatures \cite{matsui86b}. 
The idea was based on an analogy
with the well known Mott transition in condensed matter systems.  At high
densities, Debye screening in a quark-gluon plasma
reduces the range of the attractive force
between heavy quarks and antiquarks, and above some critical density
screening prevents the formation of bound states.  The larger bound
states are expected to dissolve before the smaller ones as the
temperature of the system increases.  The $\psi'$ and $\chi_c$ states
are thus expected to become unbound just above $T_c$, while
 the smaller $\psi$
state may only dissolve above $\approx 1.2 T_c$.  Heavier $b\bar{b}$
states offer the same features as $c\bar{c}$ states, but require much
shorter screening lengths to dissolve \cite{karsch91a}.  The
$\Upsilon(b\bar{b})$ state may dissolve only around 2.5 $T_c$, while
the larger excited $\Upsilon'$ could also dissolve near $T_c$.

At SPS energies, less than one charmonium is on average produced per 
heavy-ion reaction \cite{karzeev,blaizot} -- 
here the screening argument is most compelling and
the debate has shifted towards competing hadronic processes 
\cite{comovers}: the size
of the hadronic absorption cross sections for charmonia remains a contentious 
issue. However, these cross sections are necessary to estimate the 
contribution of the hadronic phase to charmonium suppression
and unless a more solid understanding of these 
cross sections is obtained, the interpretation of the SPS charmonium data
will remain under debate.
At RHIC, several $\bar cc$ pairs may be created in one event -- thus
allowing for a coalescence of a $\bar cc$ pair which was not produced
in the same hard process. Estimates taking this effect into account
actually predict an enhancement of charmonium production in a QGP, compared
to a purely hadronic scenario \cite{schroeter}.

\section{Outlook}

Thanks to the vast amount of exciting new RHIC data, QGP theory is currently
the most active, dynamic and exciting field in theoretical nuclear physics. 
Already the current data are of such high quality that they might allow
not only the observation of a deconfined phase of matter, but also the
characterization of the QGP equation of state! In order to accomplish this
feat, a consistent picture of all phases of a heavy-ion
reaction needs to be developed, not limited to a few chosen observables.
Proper baselines for comparison need to be established as well -- the
importance of the RHIC p+p and p+A program cannot be overstated: these
experiments will serve to constrain theories and models as well as help
to generate a better understanding of initial state effects.

There are a number of 
milestones which QGP theory needs to address in the forthcoming years
to fulfill the promise the data is currently showing us:
\begin{itemize}
\item the development of reliable lattice gauge calculations at
	finite temparature {\em and} chemical potential
\item the determination of hadron properties at high densities and temperatures
\item generating a better understanding on the mechanisms of hadronization,
	both, for elementary hadron-hadron reactions as well as for
	bulk QCD matter
\item the link-up of initial state saturation models to the later-stage
	dynamics of a heavy-ion collision, including pQCD processes
\item the investigation of mechanisms for chemical equilibration, both in
	the confined as well as in the deconfined phase
\item the development of a transport theory based on QCD which treats
	both, hard as well as soft processes consistently.
\end{itemize}

\section*{Acknowledgments}
This work was supported by  RIKEN, Brookhaven
National Laboratory and DOE grants DE-FG02-96ER40945 and
DE-AC02-98CH10886.

\end{document}